\documentclass[aps,prl,twocolumn]{revtex4-2}

\usepackage{amsmath}
\usepackage{amssymb}
\usepackage{bm}
\usepackage{hyperref}

\usepackage{xcolor}
\usepackage{braket}
\usepackage{graphicx}

\usepackage{afterpage}

\usepackage[normalem]{ulem} 

\usepackage{bbold}
\usepackage{blkarray}
\usepackage{ulem}
\usepackage[final]{pdfpages}
\usepackage{pgffor}
\makeatletter
\AtBeginDocument{\let\LS@rot\@undefined}
\makeatother

\begin{document}

\title{Anyon superconductivity and plateau transitions in doped fractional quantum anomalous Hall insulators}

\author{Pavel A. Nosov}
\thanks{These authors contributed equally to this work.}
\affiliation{Department of Physics, Harvard University, Cambridge, MA 02138, USA}

\author{Zhaoyu Han}
\thanks{These authors contributed equally to this work.}
\affiliation{Department of Physics, Harvard University, Cambridge, MA 02138, USA}

\author{Eslam Khalaf}
\affiliation{Department of Physics, Harvard University, Cambridge, MA 02138, USA}

\date{\today}
\begin{abstract}
Recent experiments reported evidence of superconductivity and re-entrant integer quantum anomalous Hall (RIQAH) insulator upon doping the $\nu_e = 2/3$ fractional quantum anomalous Hall states (FQAH) in twisted MoTe${}_2$, separated by narrow resistive regions. Anyons of a FQAH generally have a finite effective mass, and when described by anyon-flux composite fermions (CF), experience statistical magnetic fields with a commensurate filling. Here, we show that most of the experimental observations can be explained by invoking the effects of disorder on the Landau-Hofstadter bands of CFs. In particular, by making minimal assumptions about the anyon energetics and dispersion, we show that doping anyons drives plateau transitions of CFs into integer quantum Hall states, which physically corresponds to either a superconductor or to a RIQAH phase. We develop a dictionary that allows us to infer the response in these phases and the critical regions from the knowledge of the response functions of the plateau transitions. In particular, this allows us to relate the superfluid stiffness of the superconductor to the polarizability of CFs. As a first step towards a quantitative understanding, we borrow results from the celebrated integer quantum Hall plateau transitions to make quantitative predictions for the critical behavior of the superfluid stiffness, longitudinal and Hall conductivity, and response to out-of-plane magnetic field, all of which agree reasonably well with the experimental observations. Our results provide strong support for anyon superconductivity being the mechanism for the observed superconductor in the vicinity of the $\nu_e = 2/3$ FQAH insulator.
\end{abstract}
\maketitle

\begin{figure}[t!]
    \centering
\includegraphics[width=0.90\linewidth]{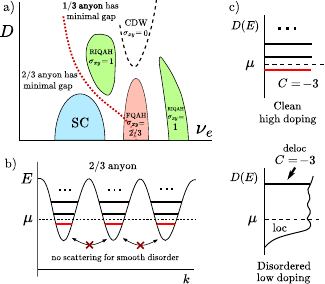}
    \caption{a) Cartoon of the finite temperature phase diagram in the vicinity of the $\nu_e=2/3$ FQAH state, as a function of doping $\nu_e$ and displacement field $D$ (after \cite{MoTe2SC}). The red dotted line separates regions where we assume that $2/3$- and $1/3$-anyons are energetically more favorable.  b) $2/3$-anyon dispersion with three minima in a Brillouin zone without disorder. Red solid lines denote filled LLs that appear due to statistical flux. Introduction of smooth potential disorder approximately preserves valley symmetry, but broadens levels within each valley. c) The density of states (with and without disorder) for the $2/3$-anyons. }
    \label{fig:diagrams}
\end{figure}

\emph{Introduction}--- Recent experiments have reported the remarkable discovery of fractional quantum anomalous Hall (FQAH) states, a zero-field analog of the fractional quantum Hall (FQH) states \cite{liuRecentDevelopmentsFractional2022,parameswaranFractionalQuantumHall2013,BergholtzReview2013,neupertFractionalQuantumHall2011,shengFractionalQuantumHall2011,regnaultFractionalChernInsulator2011,qi_generic_2011,parameswaranFractionalChernInsulators2012,wuBlochModelWave2013,kourtisFractionalChernInsulators2014, zhangNearlyFlatChern2019, tarnopolskyOriginMagicAngles2019,wu2019topological,ledwithFractionalChernInsulator2020a,wangChiralApproximationTwisted2021a, repellinFerromagnetismNarrowBands2020, meraEngineeringGeometricallyFlat2021,meraKahlerGeometryChern2021,ledwithFamilyIdealChern2022,vortexability,lu2025electromagnetic,Tianhong2025,abouelkomsan2023quantum, abouelkomsanParticleHoleDualityEmergent2020a}, in twisted MoTe${}_2$ (t-MoTe${}_2$)~\cite{cai2023signatures, zeng2023thermodynamic, park2023observation, xu2023observation} and in pentalayer rhombohedral graphene aligned to an hBN substrate~\cite{lu2024fractional}. Subsequent experiments have reported the even more surprising discovery of superconductivity in these two systems. First, Ref.~\cite{RhombohedralSC} reported the observation of superconductivity in moir\'eless rhombohedral graphene, which develops in the quarter metal phase where valley and spin are likely fully polarized and time-reversal is strongly broken. More recently, Ref.~\cite{MoTe2SC} reported the observation of superconductivity in a high-quality t-MoTe${}_2$ sample where previous optical measurements suggest similar full flavor polarization~\cite{cai2023signatures,anderson2024trion}. Remarkably, this phase emerges close to a FQAH state at filling fraction $\nu_e = 2/3$ of the moir\'e band, separated by a narrow resistive region on the doping axis. The vicinity of the $\nu_e = 2/3$ is also marked by several re-entrant integer quantum anomalous Hall (RIQAH) states realized by tuning either the doping $\nu_e$ or the vertical displacement field $D$. A sketch of the finite temperature experimental phase diagram, adapted from Ref.~\cite{MoTe2SC}, as a function of $\nu_e$ and $D$ is given in Fig.~\ref{fig:diagrams}a.

\begin{figure*}[t!]
    \centering
\includegraphics[width=0.78\linewidth]{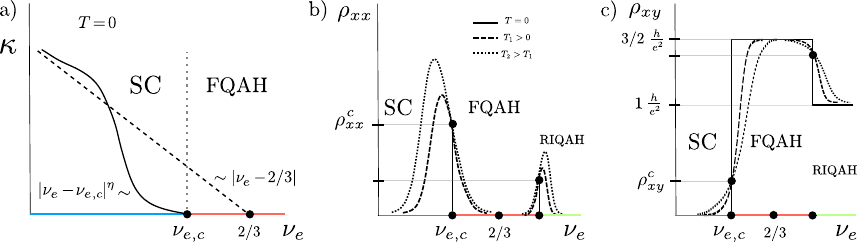}
    \caption{a) Schematic behavior of the superfluid stiffness $\kappa$ at $T=0$ and $D=0$ as a function of $\nu_e$. b,c) Schematic behavior of the (b) longitudinal resistivity $\rho_{xx}$ and (c) Hall resistivity $\rho_{xy}$ as a function of doping, at several different temperatures. The temperature dependence of $\rho_{xx}, \rho_{xy}$ is more pronounced in the SC and RIQAH states as compared to the $2/3$ plateau itself because their characteristic energy scales are governed by the doped anyon density and, thus, are much smaller than the parent FQAH gap. On the SC side, this asymmetry is further amplified by small superfluid stiffness at small doping.}
    \label{fig:StiffnessResistivity}
\end{figure*}

The emergence of superconductivity upon lightly doping an FQAH suggests that it is possibly associated with the doped anyons. The idea of anyon superconductivity (SC) dates back to classic works of Laughlin ~\cite{LaughlinAnyonSC, laughlin1988, FetterHannaLaughlin}, which suggested that a gas of itinerant anyons has a natural instability to SC. This arises from statistical interactions among anyons, which can be described by composite fermions (CF) coupled to a Chern-Simons (CS) gauge field. The CFs can be thought of as anyon-flux composites subject to a CS statistical flux density. Within a mean-field treatment, we can assume a uniform CS flux leading to Landau levels (LLs) of CFs. Due to the CS flux attachment relation, the density of the CFs is tied to the CS flux density, leading to a fixed LL filling. For certain types of anyons, this filling is an integer, thus can lead to an integer quantum Hall (IQH) state of CFs. Adding external magnetic flux to this state amounts to moving CFs across a LL gap, which cost a finite energy. This indicates the resulting electron state is a superconductor (SC)~\cite{LaughlinAnyonSC, FetterHannaLaughlin, 
laughlin1988, WilczekWittenHalperinAnyonSC,PhysRevLett.63.903,divic2024anyon,PhysRevB.39.11413,PhysRevB.42.342,PhysRevB.41.240,doi:10.1142/S0217979291001607,PhysRevB.41.11101,divic2024anyon}.

Recently, Ref.~\cite{ShiSenthil} has revived this idea in the context of a doped FQAH and presented a parton description, resulting in an effective Chern-Simons theory that clarified important subtleties and distinctions compared to earlier construction. 
A crucial distinction between the doped anyons of a FQAH realized in a lattice system compared to those of FQH in uniform field LLs is the absence of continuous magnetic translation symmetry (MTS). This allows for a finite dispersion for anyons even when the single-particle dispersion is flat (see recent works~\cite{pichler2025microscopic,schleith2025anyon,gonccalves2025spinless} for examples). 
Importantly, while flat bands are needed for stabilizing FQAH phases, uniform quantum geometry is not required; recent work has identified a condition, called vortexability~\cite{roy2014band, jackson2015geometric, claassen2015positionmomentum, lee2017band,  ledwithFractionalChernInsulator2020a, ozawa2021relations, meraKahlerGeometryChern2021, meraEngineeringGeometricallyFlat2021, wang2021exact, ledwith2023vortexabilitya}, that is sufficient to guarantee a Laughlin FQAH groundstate for short-range interactions, irrespective of the non-uniformity of quantum geometry. In fact, recent calculations suggest sizeable Berry curvature fluctuations in t-MoTe${}_2$ flat bands~\cite{dong2023composite, TMDAharonocCasher}.


\emph{Qualitative picture}--- Our results are summarized in Figs.~\ref{fig:diagrams} and \ref{fig:StiffnessResistivity}. We consider doping a FQAH at fixed disorder strength. In terms of the CFs, doping induces an effective magnetic field, such that the only relevant energy scale at small doping is an effective cyclotron frequency, $\hbar \omega_c \propto |\delta\nu|$ where $\delta \nu \equiv \nu_e - 2/3$ is the doping concentration. To account for disorder, we introduce the scattering rate $1/\tau$, which can be combined with $\omega_c$ to form a dimensionless measure of the disorder strength, $\omega_c \tau$. This parameter increases as a function of doping at a fixed $\tau$, thus it is convenient to think of increasing doping as decreasing disorder strength.

There are two types of anyons with charge $2/3$ and $1/3$ which have self-statistical angles $2\pi/3$ and $-\pi/3$, respectively. Since the parent FQAH state can be viewed as already having one $2/3$ anyon or two $1/3$ anyons per unit cell, the doped anyons feel a background effective magnetic field that triply folds the Brillouin zone (BZ), and a discrete MTS enforces the dispersion to be three-fold degenerate therein~\cite{ShiSenthil, gonccalves2025spinless, musser2024fractionalization} (illustrated in Fig.~\ref{fig:diagrams}b). Upon doping $2/3$ or $1/3$ anyons, the additional statistical flux will reorganize the anyon bands into Landau-Hofstadter sub-bands, and the rigid relation between charge and flux implies an effective filling fraction $\nu_\text{eff} = -3$ or $3/2$ of these \emph{sub-bands} for the corresponding CFs, respectively. At very small doping $\delta \nu \ll 1$, we can understand the Landau-Hofstadter bands as LLs forming in the 3-fold minima (valleys) of the anyon dispersion, leading to 3-fold degenerate LLs (Fig.~\ref{fig:diagrams}b). 


The effects of the disorder can be captured, at least on a qualitative level, by ignoring interactions between CFs and viewing the problem as that of LLs for single particles. We will discuss the validity of this assumption and possible caveats later. Borrowing results from IQH plateaus, we know that any amount of disorder will localize all states within a given LL except for a single extended state. For sufficiently strong disorder, such states will be pushed up in energy, leading to a trivial state for CFs. Such state will precisely correspond to the parent FQAH. Doping charge effectively amounts to reducing disorder strength, which results in the extended states going down in energy until they cross the chemical potential, leading to a plateau transition. The resulting state depends on the kind of doped anyon and on the number of extended states crossing the Fermi energy.

A direct transition to a SC is realized if the doped charge goes in as a $2/3$ anyon \emph{and} all three delocalized states are approximately degenerate and cross the Fermi energy simultaneously. This is the case if the minima of the $2/3$ anyon dispersion are relatively deep and if the disorder is sufficiently smooth such that it does not scatter between these minima (cf.~Fig.~\ref{fig:diagrams}b). This suggests that increasing the disorder or the effective mass of the anyon can lead to a splitting of this transition into 3 transitions. This may explain the shoulder feature observed in the resistive transition region in experiment \cite{MoTe2SC}. On the other hand, a transition to a RIQAH is realized if the doped charge is a $1/3$ anyon and only one delocalized state is occupied. This suggests that $1/3$ anyons have a shallower dispersion and non-degenerate Hofstadter bands at the relevant doping. Physically, this implies a smaller characteristic energy scale for the RIQAH state compared to the SC, which is consistent with the experiment~\cite{MoTe2SC}. The resistive regions between the FQAH and either phase can be understood as IQH plateau transitions. 

The phase diagram as a function of displacement field $D$ can be attributed to changes in the anyon dispersion with $D$ due to changes in the flat band quantum geometry. For instance, the observation that the SC (IQAH) phase boundary moves away from (towards) $\nu_e= 2/3$ suggests the anyon effective mass (which controls the critical $\omega_c \tau$) decreases (increases) with $D$ for the $2/3$($1/3$)-anyon. Furthermore, the experiment suggests there is a phase boundary in the $\nu_e$-$D$ phase diagram where the cheapest charged excitation changes from $2/3$- to $1/3$-anyons. Energetically favored $2/3$-anyons indicate a tendency of $1/3$-anyons to pair as suggested by recent numerics~\cite{YangAnyonPairing, JainAnyonPairing}. The prominence of this pairing on doping $2/3 + \delta\nu$ compared to $2/3 - \delta\nu$ is consistent with earlier results studying anyon energetics in Chern-Simons-Ginzburg-Landau theory of composite bosons~\cite{tafelmayer1993topological,PhysRevB.85.241307}.

\emph{Effective field theory}--- Our analysis builds formally on an effective Lagrangian describing anyons near a $\nu_e=2/3$ FQAH state~\cite{ShiSenthil} in a single Chern band:
\begin{align}\label{eq: L}
    \mathcal{L} =& \mathcal{L} [\psi_{2/3};a] + \mathcal{L} [\psi_{1/3};b]+\mathcal{L}_\text{CS}+ \mathcal{L}_\text{dis}+\mathcal{L}_\text{int} \\
     \mathcal{L}_\text{CS}  =& \frac{1}{2\pi} \left(b-a\right) \mathrm{d} A +\frac{1}{4\pi} A\mathrm{d} A + A_0 -b_0 \nonumber\\
 & \ \ \ \ \ \ - \frac{1}{4\pi}\begin{pmatrix}
a & b
\end{pmatrix} \begin{bmatrix}
-2 & 1 \\
1 & 1
\end{bmatrix} \begin{pmatrix}
da \\ db
\end{pmatrix} \label{eq: LCS}
\end{align}
where $a\mathrm{d}b = \epsilon^{\mu\nu\eta}a_\mu \partial_\nu b_\eta$, $a$ and $b$ are emergent $U(1)$ gauge fields, and $A$ is the external gauge field. $A_0$ and $b_0$ are time components of $A$ and $b$, and we have set the crystal unit cell area $\mathcal{A}_\text{UC}=1$. $\psi_{2/3}$ and $\psi_{1/3}$ are two distinct CF fields, which represent, as we show below, the two types of anyons with charges $2/3$ and $1/3$. Finally,  $\mathcal{L}_\text{int}$ describes electron-electron interactions, while $\mathcal{L}_\text{dis}$  encodes a disorder potential coupled to the electron density. Our analysis primarily focuses on smooth, long-wavelength disorder, as is relevant for t-MoTe${}_2$, where the concentration of local impurities is estimated to be several orders of magnitude smaller than the electron density \cite{park2025observation}. This suggests that the dominant disorder could likely originate from twist-angle inhomogeneity. The short-wavelength disorder has profound effects on the transition, as will be discussed in the {\it{``Conclusions"}} section. A parton derivation of this action can be found in Ref.~\cite{ShiSenthil} and is reproduced in Supplemental Materials (SM) for completeness. The flux attachment scheme implied by this action is summarized in the End Matter.

Depending on the details in the microscopic Hamiltonian, the two anyons may have different activation gaps such that the doped charge may enter the system as either type of the anyons. Below we will show the two cases that naturally lead to the observed daughter phases in Fig.~\ref{fig:diagrams}.

\emph{Doping the $2/3$-anyons---} When the doped particles are $2/3$-anyons, the corresponding CFs, $\psi_{2/3}$, are at effective filling fraction $\nu_\text{eff} = -3$ of their self statistical flux. Therefore, the most natural state is that $\psi_{2/3}$ form a IQH state with $C=-3$ described by $-\frac{3}{4\pi} ada$. The topological response theory describing the resulting state is: $
    \mathcal{L} =   \frac{2 }{2\pi} b \mathrm{d} A +\frac{2}{4\pi} A\mathrm{d} A  $.
The fact that a charge-$2$ current $J^\mu = \frac{2}{2\pi}\epsilon^{\mu\nu\eta} \partial_\nu b_\eta$ tends to screen the external magnetic field implies that this is a SC. In SM, we further deduce that the chiral central charge of the SC is $c_-=-2$ and the pairing angular momentum (in the disorder-free limit) $L=3\mathcal{S}$ mod $3$ (assuming $C_3$ invariance) is associated with the fractional discrete shift $\mathcal{S}$ of the parent FQAH state~\cite{qslx-ybf6,8bpm-qbzp}, the latter of which is a crystalline topological invariant determining the fractional charge bound to a unit lattice disclination and, dually, the fractional angular momentum bound to a unit magnetic flux. This leads to a concrete prediction relating the fractional disclination charge to the pairing symmetry of the order parameter.

The phase stiffness of SC in 2D has the unit of energy and limits the critical temperature $T_c$ of the Berezinskii-Kosterlitz-Thouless transition out of strongly-coupled SC phases in 2D (see Ref.~\cite{PhysRevB.41.11101} for an discussion pertinent to anyon SC). In SM, we derive an expression for the stiffness in terms of the polarization tensor $\Pi^{\mu\nu}_{2/3} = \langle J^\mu J^\nu\rangle_{2/3}$ of $\psi_{2/3}$ by integrating out small fluctuations in the gauge field. We find that the inverse stiffness $\kappa^{-1}$ is simply related to the polarizability $\chi_{2/3}$ of the $\psi_{2/3}$ state:
\begin{align}\label{eq:chi_kappa}
 \kappa^{-1} = \frac{\chi_{2/3}}{(2\pi)^2}\equiv \lim_{\bm{{q}}\rightarrow 0 }\frac{\Pi_{2/3}^{00} }{\bm{q}^2} \equiv  \lim_{\omega\rightarrow 0} \lim_{\bm{q}\rightarrow 0} \frac{\Pi_{2/3}^{ii}}{\omega^2} \;.
\end{align}

Deep in the SC phase, we expect $\omega_c\tau$ to be large (disorder effects are weak) since $\omega_c = \pi |\delta\nu|/m_{2/3}$. Previous work has evaluated $\chi$ in this case~\cite{burmistrov2002two}. Building on this result, we find
\begin{equation}
    \kappa \approx \frac{\omega_c}{ 3\pi} \left(1+\frac{\pi}{6\omega_c \tau}+\mathcal{O}\left((\omega_c\tau)^{-2}\right)\right)\;.\label{eq:kappa_omega_c}
\end{equation}
The positive sign of the leading-order correction stems from the fact that the fermion polarizability in a lowest Landau level is reduced by weak disorder.
For our problem, this gives rise to a weak {\it enhancement} of $\kappa$ compared to the clean limit result. Note that this is very different from conventional BCS superconductors, where weak disorder only suppresses coherence length and $\kappa$. For smaller doping, where $\omega_c\tau\sim 1$, higher-order corrections to Eq.~\eqref{eq:kappa_omega_c} must be taken into account.


\emph{Transition region}--- As one approaches the transition region from the superconducting side, the density (and consequently the statistical flux density) decreases, indicating strong mixing between LLs within each valley due to potential scattering. In the limit $\omega_c\tau\ll 1$, the LLs significantly overlap, and all eigenstates become localized except for a discrete set of isolated extended (critical) states. These delocalized states originate near the centers of their respective Landau levels and gradually levitate relative to the Fermi level as the density continues to decrease. Thus, as long as the Fermi level stays within the mobility gap above the lowest delocalized level, the superconducting state is stable, and its stiffness can be extracted from the polarizability of the IQH state formed by $\psi_{2/3}$, which is given by $\chi\propto  m_{2/3} \xi^{2}$~\cite{Viehweger1991,Feigelman_2018} (i.e. the inverse of the level spacing within the localization volume), where $\xi$ is the localization length. A derivation of this result from Mott's argument is included in the SM. The dependence on $\xi$ (and thus the associated dynamical scaling exponent $z=2$, i.e. $\chi\sim \xi^z$) can be traced to the scaling of the dipole matrix elements $\langle m|\hat x_i|n \rangle \sim \xi$ in the current-current correlator $\Pi_{2/3}^{ii}$ and the fact that the system remains compressible at the transition with short-ranged interactions. 


At some critical filling fraction $\nu_{e,c}$, the chemical potential coincides with the lowest delocalized state, signaling an IQH–insulator transition for $\psi_{2/3}$. Near this point, the localization length diverges as $\xi\sim 1/|\nu_{e}-\nu_{e,c}|^{\nu}$, where $\nu$ is the correlation length exponent characterizing the transition. As a result, the superfluid stiffness, which is inversely proportional to the polarizability Eq.~\eqref{eq:chi_kappa}, vanishes at this transition, and we find the following critical behavior for the zero-temperature stiffness: 
\begin{equation}
    \kappa(T{=}0) \propto |\nu_{e}-\nu_{e,c}|^{\eta}, \quad \eta=2\nu\;.
\end{equation}
On general grounds, we expect that $\nu{\geq} 1$ (and thus $\eta{\geq} 2$) due to the Harris criterion \cite{Harris_1974}.  In particular, for the regular IQH–insulator transition, experiments find $\nu\approx 2.38 \pm 0.02$ \cite{PhysRevLett.94.206807,PhysRevLett.102.216801}. However, we emphasize that this value is merely suggestive as the actual critical exponent could be affected by the interplay between gauge fluctuations and disorder at the transition. Nevertheless, our simple analysis indicates that the stiffness should remain relatively low in the immediate vicinity of the transition, which could lead to much stronger temperature variations of the resistivity on the superconducting side of the plateau transition (see Fig.~\ref{fig:StiffnessResistivity}(b)). We note that this behavior was indeed observed in \cite{MoTe2SC}.

The critical filling $\nu_{e,c}$ can be roughly estimated from the well-known condition~\cite{Laughlin_extended,khmelnitskii1983quantization} determining when the lowest critical state crosses the chemical potential (which in turn follows from the two-parameter scaling picture of the IQH-insulator transition and is applicable both for $\omega_c\tau\gg 1$ and $\omega_c\tau\ll 1$.). Applying this condition to our problem, we find
$|\nu_{e,c}|\approx m_{2/3}/(\pi\tau)$, which can be intuitively understood as the point at which the broadened Landau levels of CFs begin to overlap and the single-particle gap closes (see End Matter for details.)
Thus, the critical density increases with either increasing disorder strength or the anyon effective mass. This naturally explains why this transition will not be observed in very disordered samples or in FQH with continuous MTS where $m_{2/3}$ is infinite. We note that, in principle, the precise values of the critical densities cannot be fully derived from the IQH–insulator transition of the CFs alone, as the dynamical gauge fluctuations are also likely to contribute \footnote{Similar issues with the CS gauge fluctuations arise in the more conventional context of transitions between FQH plateaus (see, e.g., \cite{Pruisken1999}).}. Nevertheless, we still expect our mean field treatment to correctly capture the qualitative trends as the anyon mass and elastic scattering rate are varied.

Using our dictionary to map the transition from FQAH to SC to an IQH plateau transition of CFs also enables us to express the conductivity tensor $\hat{\sigma}_e$ for electrons in terms of the conductivity tensor $\hat{\sigma}_{2/3}$ for $\psi_{2/3}$ as
\begin{equation}
    2\pi \hat{\sigma}_e= 2 \hat{E} + 4\left[2\pi\hat{\sigma}_{2/3} + 3\hat{E}\right]^{-1}   \;,
\end{equation}
where $ (\hat{E})^{ij}=\epsilon^{ij}$ is the unit anti-symmetric tensor (see SM for the derivation). Under the assumption that $\psi_{2/3}$ is at the regular IQHI transition, the conductivity tensor then becomes $2\pi\hat{\sigma}^c_{2/3} = (3/2)(\hat{1}- \hat{E}) $ \cite{Kumar2019,PhysRevB.46.2223} (the extra factor of $3$ arises because of the valley degeneracy), and we find $ 2\pi \hat{\sigma}^c =\frac{4}{3} \hat{1} + \frac{2}{3} \hat{E}$. In this case, the longitudinal resistivity for electrons is $\rho_{xx}^c/2\pi =3/5$ and the Hall resistivity is $|\rho_{xy}^c|/2\pi =3/10$ at the SC-FQAH transition. This corresponds to a peak resistivity $\rho_{xx}^c \approx 15 {\rm k}\Omega$  in the transition region which is close to but slightly larger than the experimentally observed value \cite{MoTe2SC}. We emphasize that the actual value of $\sigma^c_{2/3,xx}$ may differ from the one obtained using the IQH plateau transition values due to the effects of interactions and gauge fluctuations.

Finally, we note that a serious analysis of the critical point itself involves the interplay of disorder, electron-electron interactions, and gauge fluctuations -- a formidable task. While we do not attempt to explore this transition in full complexity here, in the End Matter we state the problem in the framework of a modified  Pruisken's non-linear sigma model with a topological $\theta-$term~\cite{Pruisken1983,khmelnitskii1983quantization,laughlin1983anomalous}, and comment on crucial differences distinguishing our theory from regular IQHI transitions \cite{MirlinReview,Burmistrov_review}.

\begin{figure}
    \centering
    \includegraphics[width=\linewidth]{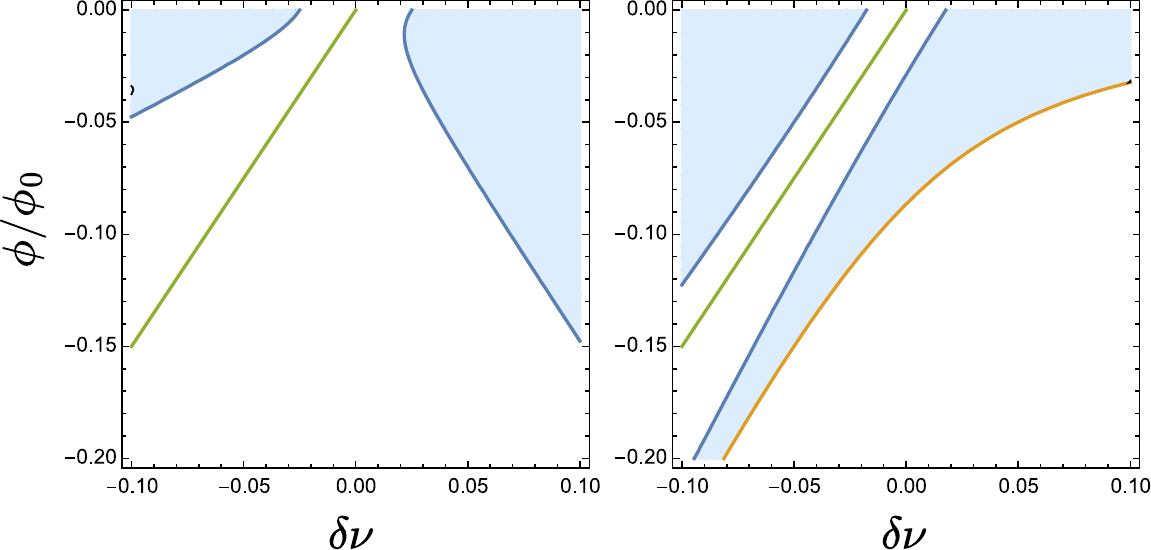}
    \caption{The boundary of the phases of interest, which are shaded in the plot. $\phi/\phi_0$ is the flux per unit cell.  (Left) the boundaries of anyon superconductor resulting from doped $2/3$ anyons. (Right) the boundaries of RIQH state resulting from doped $1/3$ anyons. The green line is defined by $\delta \nu = \frac{2}{3}\delta\nu_B$ which represents the center of FQAH state. The parameters used in these plots are $\frac{\pi \tau}{m} = 10$ for $1/3$ anyons and $\frac{\pi \tau}{m} = 40$ for $2/3$ anyons, so chosen such that the results agree reasonably with Ref.~\cite{MoTe2SC}. Note that, in principle, the masses of an anyon on the electron- or hole-doping sides may differ, but here we neglect this fact for simplicity.}
    \label{fig: wannier}
\end{figure}

\emph{Doping the $1/3$ anyon}--- We now briefly discuss the RIQAH along the same lines of our discussion of the FQAH to SC transition. When the doped particles are $1/3$ anyons, they are at effective filling fraction $\nu_\text{eff} = 3/2$ of their self statistical flux. In the clean and small doping limit, we may expect a three-fold degenerate composite Fermi liquid (CFL) corresponding to the three potential minima~\cite{ShiSenthil}. However, we expect such state to be unstable to moderate disorder. Furthermore, we note that with disorder, the transition is expected to happen at finite doping where an effective LL description in each of the potential minima will break down if the dispersion is sufficiently shallow. Instead, we should consider the Landau-Hoftadter sub-bands, the lowest few of which are generally non-degenerate. Thus, we can effectively neglect the degeneracy between the valleys under the assumption that $m_{1/3}$ is large. Then, disorder will again localize most of the states in each sub-band except one state that carries the Hall response. Since the $1/3$ anyon is at filling $3/2$, at most one delocalized state can pass the chemical potential as the disorder strength is reduced, resulting in a  $\frac{1}{4\pi} b\mathrm{d}b  $ response. The resulting effective response theory (after integrating out $b$ which set $a = A$) is $\mathcal{L} = \frac{1}{4\pi} A\mathrm{d} A $ which describes an IQH state. An analysis of the critical region similar to that for the SC transition leads to $\rho^c_{1/3,xx}/2\pi = 1/5$ and $\rho_{1/3,xy}^c/2\pi = 7/5$. This corresponds to $\rho^c_{xx,1/3} \approx 5{\rm k}\Omega$ which is very close to the value in experiment~\cite{MoTe2SC}.

\emph{Effect of magnetic field}--- Upon applying an out-of-plane magnetic field, the density of CFs as well as the effective field strength felt by them both change according to the flux attachment conditions Eq.~\ref{eq: flux attachment} and Eq.~\ref{eq: charge density}. This can lead to phase transitions out of the phases of interest by moving the Fermi level of $\psi$ across an extended state. To model this effect, we again use the levitation formula Eq.~\ref{eq:Levitation_formula} to infer the transition condition.
The resulting phase boundaries of the anyon SC and RIQAH state caused by $\psi_{2/3}$ and $\psi_{1/3}$, respectively, are shown in Fig.~\ref{fig: wannier}. We find that the choice $\frac{\pi \tau}{m} = 10$ for $1/3$ anyons and $\frac{\pi \tau}{m} = 40$ for $2/3$ anyons
approximately reproduces the experimentally observed phase boundaries~\cite{MoTe2SC}. This is consistent with our assumption of a shallower dispersion for the $1/3$ anyon.

Interestingly, we find that all the phase boundaries of the RIQAH state have a slope $\delta\nu/(\phi/\phi_0) \sim 0.6 < 1$ despite the state has $2\pi \sigma_{xy} = 1$. This violation of the Streda formula should not be surprising since the RIQAH constructed here is a {\it compressible} state and thus the resistive dip feature is not tracking a gap position. The dependence of the phase boundary on out-of-plane field seen in Fig.~\ref{fig: wannier} can be understood by noting that the plateau transitions can be induced not only by varying the filling fractions but also by tuning $\omega_c \tau$~\cite{Laughlin_extended,khmelnitskii1983quantization}.  For example, the phase boundaries of RIQAH state on the positive doping side in Fig.~\ref{fig: wannier} are predominantly determined by the change of $\omega_c\tau$, resulting in a significant deviation from the Streda formula result. Indeed, this discrepancy has been observed in Ref.~\cite{MoTe2SC}, and further compressibility measurements~\cite{yu2022correlated} may verify or falsify our theory.

\emph{Conclusions}--- We close by discussing a few predictions for future experiments. Our results suggest that away from the clean limit, 
we generally expect small lifting of the CF LL degeneracy, which would split the transition between the FQAH to SC into 3 transitions with intervening RIQAH. Possible indication of such splitting can be seen in the shoulder feature in Ref.~\cite{MoTe2SC}. We expect this splitting to be more pronounced in more disordered samples (where disorder is not strong enough to completely destroy the SC), particularly for short-range disorder with large momentum scattering, or in parameter regimes (twist angle, displacement field, etc) where the $2/3$ anyon dispersion is shallow but it is still energetically favored compared to the $1/3$ anyon. We also expect the SC to have a large tunneling gap to single-particle excitations since the excitations of the normal state are not electron-like. An interesting question, which we leave to future works, is related to the nature of the normal state above the SC. Two natural possibilities are: a gas of preformed Cooper pairs or a gas of unpaired anyons. The former will lead to transport response that is similar to a BEC superconductor~\cite{BCSBEC}, whereas the latter may have distinct transport properties which remain largely unexplored~\cite{PhysRevB.45.5504}. It would also be interesting to investigate the crossover of the SC phase to the overdoped side, which could potentially be understood as a more conventional Kohn-Luttinger 
instability in an anomalous Hall metal.

\emph{Acknowledgements}--- We acknowledge Bert Halperin and Igor Burmistrov for fruitful discussions. We are grateful to Senthil Todadri and Darius Zhengyan Shi for informing us about their manuscript~\cite{shi2025anyon} whose content overlaps with this work. Since this paper was first posted on arXiv, several related works have appeared  \cite{pichler2025microscopic, gonccalves2025spinless,huang2025apparent,guerci2025fractionalization}. E.~K. is supported by NSF MRSEC DMR-2308817 through the Center for Dynamics and Control of Materials. Z.~H. is supported by a Simons Investigator award, the
Simons Collaboration on Ultra-Quantum Matter, which
is a grant from the Simons Foundation (651440).

\bibliographystyle{apsrev4-2} 
\bibliography{bib}

         \vspace{10em}
 \onecolumngrid
        \begin{center}
             {\large \textbf{End Matter}}
         \end{center}
 \twocolumngrid
\renewcommand{\theequation}{A\arabic{equation}}
\setcounter{equation}{0}

\emph{Details on the flux attachment scheme.} The equation of motion of $a_0, b_0$ yields the flux attachment conditions:
\begin{align}
    \label{eq: flux attachment}
    \frac{1}{2\pi} \begin{pmatrix}
        \nabla\times \bm{a} \\
        \nabla \times \bm{b}
    \end{pmatrix} = & \left(\nu_{1/3} - 1\right) \begin{pmatrix}
        1/3 \\ 2/3
    \end{pmatrix} \nonumber\\ 
    & \ \ + \frac{B}{2\pi} \begin{pmatrix}
        2/3 \\ 1/3
    \end{pmatrix}  + \nu_{2/3} \begin{pmatrix}
        -1/3 \\ 1/3
    \end{pmatrix}
\end{align}
where $B$ is the external perpendicular magnetic field strength, $\nu_{a=2/3,1/3} \equiv \langle \psi^\dagger_a \psi_a\rangle $  are the average numbers of doped anyons per crystal unit cell (or simply the density, since we have set the unit cell area to $1$). Therefore, the electron density (the source of $A_0$) can be evaluated as:
\begin{align}\label{eq: charge density}
    \nu_e =  \frac{2}{3} \left(\frac{B}{2\pi} + 1 +\nu_{2/3}\right)  + \frac{1}{3} \nu_{1/3}
\end{align}
which confirms that the system is a $2\pi \sigma_{xy} = \nu_e =2/3$ state before doping. From Eq.~\ref{eq: flux attachment} and \ref{eq: charge density}, one can see that the $\psi_{2/3},\psi_{1/3}$ have effective self-statistical angles $2\pi/3, -\pi/3$ and electric charges $2/3, 1/3$, respectively. Moreover they have mutual statistical angle $2\pi/3$.  These features confirm that they describe the corresponding anyons.

\emph{Details on the Levitation formula. } The levitation formula for IQH we quoted in the main text is
\begin{equation}\label{eq:Levitation_formula}
     |\nu| = D  (n+1/2) \frac{|B_{\rm eff}|}{2\pi} \frac{1+(\omega_c\tau)^2}{(\omega_c \tau)^2 },
\end{equation}
where $\nu$ is the density of the particles, $n$ is the relevant LL index, $D$ is the degeneracy of the  LLs, $B_{\rm eff}$ is the effective magnetic field seen by them, and $\omega_c = B_\text{eff}/m$ is the cyclotron frequency that also depends on the mass of the particle. Applying to $\psi_{1/3}$ or $\psi_{2/3}$ systems, we need to plug in the effective magnetic field inferred from Eq.~\ref{eq: flux attachment}. Specifically, for the FQAH-SC transition driven by IQH transition in $\psi_{2/3}$, $\nu = \nu_{2/3} = \frac{3}{2}\nu_{e}$ and $B_\text{eff} = 2B/3  - 2\pi \nu_{2/3}/3  $, $D=3$, and $n=0$; whereas for the FQAH-RIQAH transition, $\nu = \nu_{1/3} = 3\nu_{e}$ and $B_\text{eff} = B/3 +2\pi \nu_{1/3}/3  $, $D=1$, and $n=0$.


\emph{Non-linear sigma model for a SC-FQAH transition.} --- In this section, we provide more details regarding the appropriate critical theory describing a SC-FQAH transition. In contrast to the effective description of the SC phase itself, gappless CFs cannot be simply integrated out perturbatively at the transition, and their low-energy dynamics has to be treated explicitly.  For our problem, the effective Lagrangian necessarily includes a CS gauge field leading to:
\begin{equation}\label{eq:NLSM}\begin{aligned}
    \mathcal{L}&= -\frac{3\pi}{4}\sigma_{2/3,xx}^{(0)}\operatorname{Tr}[ (\mathcal{D}_{i}  Q)^2 -iz\partial_\tau  Q -a_0 Q]\\
    &+\frac{3 \pi}{4}\sigma_{2/3,xy}^{(0)}\epsilon^{ij} \operatorname{Tr}[Q \mathcal{D}_{i}Q \mathcal{D}_{j} Q] +\mathcal{L}_{\rm CS}+\mathcal{L}_{\rm int}
    \end{aligned}
\end{equation}
where $i,j{=}x,y$, $\mathcal{D}_{\mu}Q=\partial_\mu Q-i [\delta a_\mu, Q]$ which includes covariant coupling to the gauge field fluctuations $\delta a_\mu$ on top of the mean field state, the parameter $z$ describes renormalization of the dynamical scaling exponent due to interactions~\cite{RevModPhys.66.261}, $N_r \rightarrow 0$ is the number of replicas. The constrained time bi-local matrix field $Q_{\tau,\tau'}\in U(12N_r)/U(6N_r){\times} U(6N_r)$ encodes the CF density fluctuations of $\psi_{2/3}$ (i.e. heuristically $Q\sim \bar{\psi}_{2/3}\otimes \psi_{2/3}$), with the trace and multiplications involving both the matrix structure and convolutions for time arguments. Moreover, $\sigma_{2/3,xx}^{(0)}$($\sigma_{2/3,xy}^{(0)}$) is the Drude conductivity per each valley, calculated within kinetic theory with the mean-field effective flux experienced by $\psi_{2/3}$, possibly including effects of quenched flux disorder stemming from flux attachment constraint~\cite{PhysRevB.55.15552,PhysRevB.49.16609,PhysRevX.7.031029,PhysRevB.99.235114,Kumar2022}.

In addition to the presence of a CS gauge field and a non-standard coupling to the external field $A$ (cf. Eq.~\eqref{eq: LCS}), one important difference of Eq.~\eqref{eq:NLSM} compared to the conventional (interacting) Pruisken's sigma model \cite{A_Pruisken_1995,PRUISKEN20071265,Burmistrov_review} is the existence of several approximately degenerate valleys, which allows for additional short-ranged interaction terms in $\mathcal{L}_{\rm int}$, and thus potentially more complicated renormalization flow (in a single valley case, the short-ranged interactions are known to be irrelevant at the transition \cite{DHL1996_IQH,Wang2000,BURMISTROV20111457}). If the symmetry between valleys is approximately preserved by the renormalization group flow, then this opens an intriguing possibility to analyze Eq.~\eqref{eq:NLSM} via a controlled large-$N$ expansion, where $N$ is the number of valleys. A similar approach was previously employed in the context of conventional disorder-driven SC-insulator transitions \cite{Finkelstein_LargeN}. Its extension to the present case with a topological $\theta$-term and a CS gauge field is a subject of future work. If the symmetry between the valleys is strongly broken (either by the presence of short-ranged disorder or by interaction-induced corrections), then the critical point splits into three separate transitions, with the Chern number changing by $-1$. All our considerations from the main text can be easily reproduced directly from Eq.~\eqref{eq:NLSM} provided that we assume short-ranged interactions and integrate out gauge fluctuations perturbatively. 



\section*{Supplementary material}

where we introduced a random Gaussian quenched potential $V_{\text{dis}}(\mathbf{r})$ with the following properties

$$\overline{V_{\text{dis}}(\mathbf{r})} = 0, \quad \overline{V_{\text{dis}}(\mathbf{r})V_{\text{dis}}(\mathbf{r}')} = W(\mathbf{r} - \mathbf{r}') \ . \tag{S6}$$

Here we allow $W(\mathbf{r})$ to have a finite correlation length $d$. In the limiting case of white-noise disorder, $W(\mathbf{r})$ reduces to a delta-function $W(\mathbf{r}) = \delta(\mathbf{r})/2\pi\nu\tau$. Here $\mathcal{D}_0 = m/2\pi$ is the density of states, and $\tau$ is the mean-free time between elastic collisions. If the potential was just given by a sum of uniformly-distributed delta peaks $V_{\text{dis}}(\mathbf{r}) = u \sum_i \delta(\mathbf{r} - \mathbf{r}_i)$, then the scattering rate would be given by $\tau^{-1} = mn_{\text{imp}}u^2$ (here $n_{\text{imp}}$ is the density of impurities). Note that it is independent of the anyon density because the density of states for a parabolic dispersion in 2D is a constant.

From Eq. S3 and S4, one can see that the $\psi_{2/3}, \psi_{1/3}$ have effective self-statistical angle $\theta = 2\pi/3, -\pi/3$ and electric charge $Q = 2/3, 1/3$, respectively. Moreover they have mutual statistical angle $\theta = -2\pi/3$. These features confirm that they describe the corresponding anyons.

We will consider the states that can arise with light doping and small perpendicular fields. Under this setup, the emergent gauge fields $a, b$ have approximate flux density per unit cell $-1/3, -2/3$, which suggests that the anyon bands should be considered in a magnetic unit cell that triples the crystal unit cell. Within the magnetic unit cell, all the anyon bands have dispersions with three-fold degeneracy ensured by the magnetic translation symmetry.

Depending on the details in the microscopic Hamiltonian, the two anyons may have different activation gaps such that the doped charge may enter the system as either $1/3$ or $2/3$ anyons. In the following sections, we will discussion these two cases separately.

### Doping the $2/3$-anyons: anyon superconductor

When doped particles are $2/3$ anyons, they are at effective filling fraction $\nu_{\text{eff}} = -3$ of their self statistical flux. Therefore, the most natural state is that $\psi_{2/3}$ form a IQH state with $C = -3$ described by $-\frac{3}{4\pi}ada$. The effective response theory describing the resulting state is:

$$\begin{aligned}
\mathcal{L} &= \frac{1}{2\pi}\,(b - a)\,\mathrm{d}A + \frac{1}{4\pi}A\mathrm{d}A - \frac{1}{4\pi}(a+b)\mathrm{d}(a+b) - 3\Omega_g \\
&= \frac{2}{2\pi}b\mathrm{d}A + \frac{2}{4\pi}A\mathrm{d}A - 2\Omega_g
\end{aligned} \tag{S7}$$

where in the second equality we have defined $\alpha \equiv a + b$ and integrated it out, $\Omega_g$ is a gravitational CS term that originates from the framing anomaly of the CS fields [1, 4]. (After integrating out all the dynamical gauge fields, its coefficient is the chiral central charge of the edge of the system [8], so we will explicitly keep track of this term, even though we are not going into its details.) This describes a charge-2 superconductor with current field $J = \frac{2}{2\pi} \star \mathrm{d}b$ where $\star$ is the Hodge dual operation and chiral central charge $c_- = -2$.

### Doping the $1/3$-anyons: re-entrant integer quantum Hall state

When the doped particles are $1/3$ anyons, they are at effective filling fraction $\nu_{\text{eff}} = 3/2$ of their self statistical flux. In the clean and dilute limit, there is no simple gapped state to form corresponding to this filling fraction, especially given the triple degeneracy of the effective LLs at the three valleys of the magnetic Brillioun zone (BZ). However, with finite doping concentration, we should really be thinking about the Landau-Hofstadter subbands formed in this system, the lowest few of which can be quickly non-degenerate as long as the dispersion of the anyon is shallow. Effectively, we may neglect the degeneracy between valleys under the assumption that $\psi_{1/3}$ anyons' effective mass is large. Then, disorder will again localize most of the states in each band except one state that carries the Hall response, and the levitation of the extended states may separate them such that only one of them is occupied by $\psi_{1/3}$ at the filling fraction, resulting in a $\frac{1}{4\pi}b\mathrm{d}b$ response. The resulting effective response theory of this scenario is

$$\begin{aligned}
\mathcal{L} &= \frac{1}{2\pi}\,(b - a)\,\mathrm{d}A + \frac{1}{4\pi}A\mathrm{d}A + \frac{2}{4\pi}a\mathrm{d}a - \frac{1}{2\pi}a\mathrm{d}b + \Omega_g \\
&= \frac{1}{4\pi}A\mathrm{d}A + \Omega_g
\end{aligned} \tag{S8}$$

where in the second step we integrated out $b$ which yields $a = A$. This describe an IQH state.

## PARTON DERIVATION OF THE EFFECTIVE THEORY

Here we briefly derive the effective theory, Eq. S1 used in the main text, with a parton decomposition $\psi_e = f_1 f_2 f_3$ of electrons. This derivation largely coincides with that used in Ref. [9], up to some convention difference. We choose charge assignment:

$$\mathcal{L} = \mathcal{L}[f_1; a] + \mathcal{L}[f_2; -b] + \mathcal{L}[f_3; A - a + b] \tag{S9}$$

for the emergent $U(1)$ gauge fields which glue the partons together. For the undoped case $\nu = 2/3$, we assume the flux assignment:

$$\frac{\langle \nabla \times \boldsymbol{a} \rangle}{2\pi} = -\frac{1}{3} \frac{1}{\mathcal{A}_{\mathrm{UC}}} \quad , \quad \frac{\langle \nabla \times \boldsymbol{b} \rangle}{2\pi} = -\frac{2}{3} \frac{1}{\mathcal{A}_{\mathrm{UC}}} \tag{S10}$$

such that the each parton band of a crystal BZ is reorganized into three Hofstadter sub-bands of a magnetic BZ, and two of them are filled for each type of parton according to the filling fraction. To obtain the correct response function, we assume that the $f_1, f_3$ partons form $C_1 = C_3 = 1$ Chern insulators, and $f_2$ form $C_2 = -2$ Chern insulator. This means that, in terms of the Diophantine equation of the Hoftadter problem

$$\frac{n}{n_0} = \frac{\phi}{\phi_0} C + S \tag{S11}$$

we assumed $S_1 = S_3 = 1$ and $S_2 = 2$. Since the effective response theory of such a Chern insulator is given by $\frac{C}{4\pi} A \mathrm{d}A + A_0 S + C\Omega_g$, the resulting effective theory is

$$\begin{aligned}
\mathcal{L} =& \frac{1}{4\pi} a \mathrm{d}a + a_0 \frac{1}{\mathcal{A}_{\mathrm{UC}}} + \Omega_g \\
& - \frac{2}{4\pi} b \mathrm{d}b - 2b_0 \frac{1}{\mathcal{A}_{\mathrm{UC}}} - 2\Omega_g \\
& + \frac{1}{4\pi} (A - a + b) \mathrm{d}(A - a + b) + (A_0 - a_0 + b_0) \frac{1}{\mathcal{A}_{\mathrm{UC}}} + \Omega_g \\
=& \frac{2}{4\pi} a \mathrm{d}a - \frac{1}{4\pi} b \mathrm{d}b - \frac{1}{2\pi} a \mathrm{d}b + \frac{1}{2\pi} (b - a) \mathrm{d}A + \frac{1}{4\pi} A \mathrm{d}A + \frac{1}{\mathcal{A}_{\mathrm{UC}}} (A_0 - b_0)
\end{aligned} \tag{S12}$$

Now let's consider the doped cases with excitations. These will be the parton excitations that deviate from the Chern insulator state assumed above. More specifically we will assume that the $f_3$ partons are much more gapped than $f_1$ and $f_2$, and re-name the particle excitations of $f_1$ with $\psi_{2/3}$ and the hole excitions of $f_2$ with $\psi_{1/3}$. Adding the actions of these fields to the theory above reproduces the anyon theory Eq. S1.

## THE RESPONSE THEORY

### General discussion

The physical response of the electrons can be obtained at mean-field level in terms of the resopnse of the anyons fields, $\psi$, as follows.

We will assume weak fluctuations of the emergent gauge fields, such that the response of the $\psi$ fields to the gauge fields can be effectively captured by the corresponding polarization tensor $\Pi^{\mu\nu} = \langle J^\mu J^\nu \rangle$ in a way that the effective action is quadratic in the fluctuating part of the gauge fields $\delta a, \delta b$. In the two cases of $2/3$ or $1/3$ anyons doped into the system, the effective response theory respectively reads:

$$\mathcal{L} = \frac{1}{2} \delta a_\mu \Pi^{\mu\nu}_{2/3} \delta a_\nu + \frac{3}{4\pi} \delta a \mathrm{d}\delta a - \frac{2}{2\pi} \delta a \mathrm{d}A + \frac{2}{4\pi} A \mathrm{d}A \tag{S13}$$

$$\mathcal{L} \sim \frac{1}{2} \delta b_\mu \Pi^{\mu\nu}_{1/3} \delta b_\nu - \frac{3}{2} \frac{1}{4\pi} \delta b \mathrm{d}\delta b + \frac{1}{2} \frac{1}{2\pi} \delta b \mathrm{d}A + \frac{1}{2} \frac{1}{4\pi} A \mathrm{d}A \tag{S14}$$

where we used the $\sim$ symbol in the second case since the quantization of the CS terms has been relaxed since we are only interested in the response functions here. For simplicity, we then take the temporal gauge for all the gauge

fields ($a_0 = b_0 = A_0 = 0$). Then further integrating out $\delta a$ or $\delta b$ yields the following expressions for the physical polarization tensor of electrons and thus the conductivity in the two cases, respectively:

$$2\pi\sigma_e^{ij}(\boldsymbol{q},\omega) = 2\pi\Pi_e^{ij}(\boldsymbol{q},\omega)/(\mathrm{i}\omega) = 2E - 4\mathrm{i}\omega E \cdot [2\pi\Pi_{2/3} + 3\mathrm{i}\omega E]^{-1} \cdot E$$
$$= 2E - 4E \cdot [2\pi\sigma_{2/3} + 3E]^{-1} \cdot E \tag{S15}$$

$$2\pi\sigma_e^{ij}(\boldsymbol{q},\omega) = 2\pi\Pi_e^{ij}(\boldsymbol{q},\omega)/(\mathrm{i}\omega) = \frac{1}{2}E - \frac{1}{4}E \cdot [2\pi\sigma_{1/3} - \frac{3}{2}E]^{-1} \cdot E \tag{S16}$$

where

$$E \equiv \begin{bmatrix} 0 & 1 \\ -1 & 0 \end{bmatrix} \tag{S17}$$

For example, at the critical point of the two possible plateau transitions at zero temperature discussed in the main text, we expect

$$2\pi\sigma_{2/3} = 3/2(I - E) \tag{S18}$$
$$2\pi\sigma_{1/3} = 1/2(I + E) \tag{S19}$$

where $I$ is the identity. This means that the electron conductivity in the two cases is, respectively,

$$2\pi\sigma_e = \frac{4}{3}I + \frac{2}{3}E \tag{S20}$$
$$2\pi\sigma_e = \frac{1}{10}I + \frac{7}{10}E \tag{S21}$$

which means the resistivity

$$\rho_e/(2\pi) = \frac{3}{5}I - \frac{3}{10}E \tag{S22}$$
$$\rho_e/(2\pi) = \frac{1}{5}I - \frac{7}{5}E \tag{S23}$$

**Special care for the superconducting case**

For the superconducting state, the compactness of the gauge fields is essential for obtaining the correct response theory. Technically, $\delta a$ cannot be integrated out in such a phase but requires a more careful treatment as follows.

Generically, the polarization tensor (generated by $\psi_{2/3}$) can be decomposed to:

$$\Pi_{\mu\nu}(q) = \sum_{\mathrm{i}} \varpi^{(\mathrm{i})}(q) T_{\mu\nu}^{(\mathrm{i})}(q) \tag{S24}$$

$$T_{\mu\nu}^{(\mathrm{i}=0,1,2)}(q) = v_\mu^{(\mathrm{i})} v_\nu^{(\mathrm{i})} \;,\;\; v_\mu^{(\mathrm{i})} = \frac{1}{2\pi}\epsilon_{\mathrm{i}\mu\nu}q_\nu \tag{S25}$$

$$T_{\mu\nu}^{(\mathrm{i}=3)}(q) = \frac{1}{2\pi}\mathrm{i}\epsilon_{\mu\nu\eta}q_\eta \tag{S26}$$

Or more explicitly

$$\Pi_{\mu\nu}(\boldsymbol{q},\omega) = \frac{\varpi^{(0)}}{(2\pi)^2}\begin{bmatrix} 0 & 0 & 0 \\ 0 & q_y^2 & -q_x q_y \\ 0 & -q_x q_y & q_x^2 \end{bmatrix} + \frac{\varpi^{(1)}}{(2\pi)^2}\begin{bmatrix} q_y^2 & 0 & -\omega q_y \\ 0 & 0 & 0 \\ -\omega q_y & 0 & \omega^2 \end{bmatrix} + \frac{\varpi^{(2)}}{(2\pi)^2}\begin{bmatrix} q_x^2 & -\omega q_x & 0 \\ -\omega q_x & \omega^2 & 0 \\ 0 & 0 & 0 \end{bmatrix} + \frac{\mathrm{i}\varpi^{(3)}}{2\pi}\begin{bmatrix} 0 & q_y & -q_x \\ -q_y & 0 & \omega \\ q_x & -\omega & 0 \end{bmatrix} \tag{S27}$$

In $2+1$d, any compact $U(1)$ dynamical gauge field, e.g. $a$, admits a dual representation:

$$J^\mu = \frac{1}{2\pi}\epsilon^{\mu\nu\eta}\partial_\nu a_\eta \tag{S28}$$

subject to the continuity constraint $\partial_\mu J^\mu = 0$ which can be enforced by $U(1)$ Lagrangian multiplier $\theta$. This allows us to rewrite the Lagrangian at three-momentum $q$ into the form:

$$L(q) = \frac{1}{2} \sum_{\mu=0,1,2} \varpi^{(\mu)} J_\mu J^\mu + \frac{2\pi[3+\varpi^{(3)}]}{2} \epsilon^{\mu\nu\eta} J_\mu J_\nu \frac{1}{iq_\eta} + (iq_\mu \theta - 2A_\mu) J^\mu \tag{S29}$$

Then, integrating out $J$, we obtain the final effective action

$$L(q) = \frac{\kappa^{\mu\nu}}{2} (iq\theta - 2A)_\mu (iq\theta - 2A)_\nu \tag{S30}$$

where $\kappa$ is the inverse of the kernel of $J$.

For the case where $\varpi^{(3)} = -3$ cancels the native CS term, we find the action is exactly the effective action of a charge-$2e$ superconductor. From its expression we can extract the dictionary between the compressibility $\kappa^t$ and stiffness $\kappa$ (assuming rotational invariance) for the electrons and the polarization tensor of the $\psi_{2/3}$ anyons:

$$1/\kappa^0 = \varpi^{(0)} = \lim_{\boldsymbol{q}\to 0} \lim_{\omega\to 0} \frac{(2\pi)^2}{\boldsymbol{q}^2} (\Pi_{11} + \Pi_{22}) \tag{S31}$$

$$1/\kappa = \varpi^{(i)} = \lim_{\boldsymbol{q}\to 0} \frac{(2\pi)^2}{\boldsymbol{q}^2} \Pi_{00} = \lim_{\omega\to 0} \lim_{\boldsymbol{q}\to 0} \frac{(2\pi)^2}{\omega^2} \Pi_{ii} = \chi \tag{S32}$$

where $\chi$ is, by definition, the polarizability of $\psi$. Note that the order of limits matters in the above expressions.

## REVIEW ON DISORDERED ELECTRONS IN LANDAU LEVELS

In this section we review some known results for disordered electrons in landau levels.

### The density of states

The density of states within self-consistent Born approximation (SCBA) and under the assumption $\omega_c \tau \ll 1$ ($\tau$ is the scattering time and $\omega_c = B/m$ is the cyclotron frequency) reads as

$$D(\mathcal{E}) = \frac{m}{2\pi} \left( 1 + 2\sum_{k=1}(-1)^k e^{-\frac{\pi k}{\omega_c \tau}} \cos\left(\frac{2\pi k \mathcal{E}}{\omega_c}\right) \right) \tag{S33}$$

Here $k$ labels different harmonics. This expression is expected to work well in the strongly disordered limit $\omega_c \tau \ll 1$, and not too close to the bottom of the band. The lowest order approximation in $\mathcal{O}(\exp(-1/\omega_c\tau))$ is

$$D(\mathcal{E}) \approx \frac{m}{2\pi} \tag{S34}$$

In the opposite limit $\omega_c \tau \gg 1$, we find

$$D(\mathcal{E}) = \sum_{n=0} \frac{B\Gamma_n(\mathcal{E})}{\pi^2 \Gamma}, \qquad \Gamma_n(\mathcal{E}) = \text{Re}\sqrt{(\mathcal{E} - \omega_c(n+1/2))^2 - \Gamma^2}, \qquad \Gamma = \sqrt{\frac{2\omega_c}{\pi\tau}} \ll \omega_c \tag{S35}$$

which is just a collection of semicircular broadened (but still non-overlapping) Landau levels. We note that the square root spectral edge is just an artifact of SCBA: in principle, there is always a Lifshitz tail leaking into the gap.

### The position of the delocalized states

We follow the two-parameter scaling argument to obtain the critical density at which the delocalized states crosses the Fermi level. It was suggested that the conductivities should scale as [5–7]

$$\sigma_{xx}^{(0)} = \frac{\sigma_0}{1 + (\omega_c \tau)^2}$$

$$\sigma_{xy}^{(0)} = \frac{\sigma_0 \omega_c \tau}{1 + (\omega_c \tau)^2} \tag{S36}$$

where $\sigma_0 = \rho\tau/m$ is the bare Drude conductivity at $B = 0$. The condition for the $n$-th extended state being at the Fermi level is

$$\sigma_{xy}^{(0)} = \left(n + \frac{1}{2}\right)\frac{e^2}{h} \tag{S37}$$

This means that the critical density needs to satisfy

$$\rho_c^{(n)} = \frac{B}{2\pi}\left(n + \frac{1}{2}\right)\frac{1 + (\omega_c\tau)^2}{(\omega_c\tau)^2} \tag{S38}$$

In both the weak ($\omega_c\tau \gg 1$) and strong ($\omega_c\tau \ll 1$) limits, this expression reproduces asymptotically exact results; for intermediate regime, the result can be systematically improved using the self-consistent Born approximation, which we do not include in this work.

### The Mott's law for conductivity

Here we review the derivation celebrated Mott's law on conductivity in Anderson insulators, which applies to the cases where the Fermi level in within the localized states. The goal is essentially to evaluate the current-current correlation:

$$\Pi^{ii}(\boldsymbol{q} = 0, \omega) = \langle J^i(\omega)J^i(-\omega)\rangle = \sum_{\epsilon_n < \mu, \epsilon_m > \mu} |\langle\psi_n|J^i|\psi_m\rangle|^2 \left[\frac{1}{\omega - \epsilon_{mn} + i0^+} - \frac{1}{\omega + \epsilon_{mn} + i0^+}\right] \tag{S39}$$

where $n, m$ are single particle eigenstate labels and $\mu$ is the chemical potential, and $J$ is the total current operator. We will make the following assumptions to approximately evaluate this sum:

- Every state $\psi_n$ is localized within a localization length $\xi_n \sim \xi(\epsilon_n)$, which is determined by the energy. The corresponding characteristic energy scale (typical level splitting) is $\delta_n \equiv 1/\mathcal{D}_0(\epsilon_n)\xi_n^d$ where $\mathcal{D}_0$ is the density of state and $d$ is the spatial dimension.

- The overlap integral can be approximated as

$$|\langle\psi_n|J^i|\psi_m\rangle| \approx |\boldsymbol{R}_i|\bar\delta e^{-R/\bar\xi} \tag{S40}$$

  where $\bar\delta = \sqrt{\delta_n\delta_m}$, $\bar\xi = \frac{2\xi_n\xi_m}{\xi_n+\xi_m}$, and $R \sim |\boldsymbol{R}_i|\sqrt{d}$ is the separation between the centers of $\psi_n$ and $\psi_m$.

- The sum can be approximated with a integral over the two energies and the separation between the two states

$$\sum_{\epsilon_n < \mu, \epsilon_m > \mu} \to \int_{\epsilon_1 < \mu}\int_{\epsilon_2 > \mu}\int_{\boldsymbol{R}} P(\epsilon_1, \epsilon_2, \boldsymbol{R}) \tag{S41}$$

  where $P$ is the joint probability for given energies and the relative distance of the coordinates of two localized states.

- There is level repulsion if two localized states are too close, i.e.

$$P(\epsilon_1, \epsilon_2, \boldsymbol{R}) \approx \begin{cases} 0 & \epsilon_{21} < \bar\delta e^{-R/\bar\xi} \\ \mathcal{D}_0(\epsilon_1)\mathcal{D}_0(\epsilon_2) & \text{otherwise} \end{cases} \tag{S42}$$

Putting everything together, we have the expression

$$\Pi^{ii}(\boldsymbol{q} = 0, \omega) \sim \int_{\epsilon_1 < \mu}\mathcal{D}_0(\epsilon_1)\int_{\epsilon_2 > \mu}\mathcal{D}_0(\epsilon_2)\int_{R^*}^{\infty} R^{d-1}\mathrm{d}R\frac{\epsilon_{21}R^2\bar\delta^2 e^{-2R/\bar\xi}}{(\omega + i\delta)^2 - \epsilon_{21}^2} \tag{S43}$$

where $R^* = \max(\bar\xi\ln\frac{\bar\delta}{\epsilon_{21}}, 0)$. Evaluating the integral over $R$, we obtain

$$\Pi^{ii}(\boldsymbol{q} = 0, \omega) \sim \int_{\epsilon_1 < \mu}\mathcal{D}_0(\epsilon_1)\int_{\epsilon_2 > \mu}\mathcal{D}_0(\epsilon_2)\frac{\epsilon_{21}\bar\xi^{d+2}\bar\delta^2\Gamma(d+2, 2R^*/\bar\xi)}{(\omega + i0^+)^2 - \epsilon_{21}^2} \tag{S44}$$

The imaginary part of this integral for small $\omega$ is straightforward to evaluate. Using $\frac{1}{x \pm \mathrm{i}0^+} = P\frac{1}{x} \mp \mathrm{i}\pi\delta(x)$, we obtain:

$$\mathrm{Im}[\Pi^{00}(\omega)] \sim \mathcal{D}_0^2(\mu)|\omega|\xi^{d+2}(\mu)\bar{\delta}^2\Gamma(d+2, 2\ln\frac{\bar{\delta}}{|\omega|}) \sim \mathrm{i}\mathcal{D}_0^2\omega^3\xi^{d+2}\left[\ln\frac{\bar{\delta}}{|\omega|}\right]^{d+1} \tag{S45}$$

where all the quantities are evaluated at the Fermi energy. This reproduces the Mott's law

$$\mathrm{Re}[\sigma(\omega)] = \mathrm{Im}[\Pi^{00}(\omega)]/(\mathrm{i}\omega) \sim \omega^2[\ln|\omega|]^3\xi^{d+2}\mathcal{D}_0^2 \tag{S46}$$

The real part of the integral for small $\omega$ needs a bit more care, for which we use the identity

$$P\int_0^\infty \frac{f(x)}{x^2 - y^2}\mathrm{d}x = \sum_{n=0}^\infty y^{2n}\int_0^\infty \frac{f(x)}{x^{2n+2}}\mathrm{d}x \tag{S47}$$

so that the small $\omega$ expansion to the leading nontrivial order is

$$\mathrm{Re}[\Pi^{ii}(\omega)] \sim \omega^2 \int_{\epsilon_1 < \mu} \mathcal{D}_0(\epsilon_1) \int_{\epsilon_2 > \mu} \mathcal{D}_0(\epsilon_2)\bar{\epsilon}^{d+2}\bar{\delta}^2\Gamma(d+2, 2R^*/\bar{\xi})/\epsilon_{21}^3 \tag{S48}$$

The above integral is dominated by the contribution from the small $\epsilon_{21} \lesssim \bar{\delta}$ region where both $\epsilon_{1,2}$ are close to fermi level since $\Gamma(d+2, 2\ln\frac{\bar{\delta}}{\epsilon_{21}})/\epsilon_{21}^3 \sim \epsilon_{21}^{-1}[\ln\frac{\bar{\delta}}{\epsilon_{21}}]^{d+1}$ diverges for small $\epsilon_{21}$. Therefore, we approximately have

$$\mathrm{Re}[\Pi^{ii}(\omega)] \sim \mathcal{D}_0^2\omega^2\xi^{d+2}\bar{\delta} \tag{S49}$$

where all the quantities are evaluated at the fermi energy. This implies a finite polarizability [3]

$$\mathrm{Im}[\sigma(\omega)] = \mathrm{Re}[\Pi^{00}(\omega)]/(\mathrm{i}\omega) \sim \omega\xi^2\mathcal{D}_0 \implies \chi \sim \xi^2\mathcal{D}_0 \tag{S50}$$

### The polarizability in the weak disorder limit

In the weak disorder limit, the chemical potential still lies in the gap between LLs at integer fillings (neglecting the effects of exponentially small Liftshitz tail). The polarizability in this case with $n$ filled LLs is evaluated [2] as

$$\chi = \frac{n}{4\pi\omega_c}\left(1 - \frac{\pi}{6\omega_c\tau}\right)\boldsymbol{q}^2 \tag{S51}$$

## EFFECTS OF PERPENDICULAR MAGNETIC FIELD

In this section we use the results of the previous sections to discuss the stability of the anyon superconductor, as well as the response of re-entrant IQH state, against applied magnetic field.

### Anyon superconductor

Based on Eq. S3, we see that the effective magnetic field seen by the low-energy $\psi_{2/3}$ particles in the valleys is

$$\frac{B_{\mathrm{eff}}}{2\pi} \equiv \frac{\nabla \times \boldsymbol{a}}{2\pi} + \frac{1}{3}\frac{1}{\mathcal{A}_{\mathrm{UC}}} = \frac{2}{3}\frac{B}{2\pi} - \frac{1}{3}\rho_{2/3} \tag{S52}$$

Therefore the effective cyclotron frequency

$$\omega_c = \frac{2\pi}{m_{2/3}}\left|\frac{1}{3}\rho_{2/3} - \frac{2}{3}\frac{B}{2\pi}\right| \tag{S53}$$

where $m_{2/3}$ is the effective mass of $\psi_{2/3}$. On the other hand, the density is given by deviation the filling fraction of electrons from $\nu = 2/3$:

$$\rho_{2/3}\mathcal{A}_{\mathrm{UC}} = \frac{3}{2}\delta\nu - \delta\nu_B \tag{S54}$$

where $\delta\nu_B \equiv \frac{B\mathcal{A}_{\mathrm{UC}}}{2\pi}$ is the external flux 'filling fraction' per unit cell. For convenience, we will parametrize

$$\omega_c\tau \equiv \tilde{\tau}\,|\delta\nu - 2\delta\nu_B| \tag{S55}$$

where $\tilde{\tau} \equiv \frac{\pi\tau}{m_{2/3}\mathcal{A}_{\mathrm{UC}}}$ is a dimensionless measure of the disorder strength.

We are now ready to discuss the stability of anyon superconductor in the presence of magnetic field. The transition point out of the superconducting regime, by our theory, is determined by when the 0-th (or 1-st) de-localized states in the three valleys simultaneously cross the Fermi level. Using Eq. S38, we find these two boundaries should be determined by setting $n = 0, 1$ in the below equation

$$-\frac{1}{3}\rho_{2/3} = \frac{B_{\mathrm{eff}}}{2\pi}\left(n + \frac{1}{2}\right)\frac{1 + (\omega_c\tau)^2}{(\omega_c\tau)^2}$$

$$\implies \tilde{\tau}^2\left(\delta\nu - \frac{2}{3}\delta\nu_B\right)(\delta\nu - 2\delta\nu_B) = \left(n + \frac{1}{2}\right)\left[1 + (\delta\nu - 2\delta\nu_B)^2\tilde{\tau}^2\right] \tag{S56}$$

It should be noted that, in reality, the FQAH state spontaneously breaks the time reversal symmetry and we have implicitly assumed that the $B = 0$, $\nu = 2/3$ state has $2\pi\sigma_H = 2/3$. A *positive* $B$ thus flips the Chern number of the parent state, and a time-reversed discussion should be applied in that case. Therefore, the current discussion is only relevant for $B < 0$.

### Re-entrant integer quantum Hall state

The scenario of the RIQH state is more subtle since our theory suggests that the dispersion of $\psi_{1/3}$ may be quite shallow such that there is no LL degeneracy among the valleys. A crude approximation is to neglect the valleys caused by the discrete magnetic translation symmetry and apply the above analysis again, as we are doing below.

Based on Eq. S3, we see that the effective magnetic field seen by the low-energy $\psi_{1/3}$ particles is

$$\frac{B_{\mathrm{eff}}}{2\pi} \equiv \frac{\nabla \times \boldsymbol{b}}{2\pi} + \frac{2}{3}\frac{1}{\mathcal{A}_{\mathrm{UC}}} = \frac{1}{3}\frac{B}{2\pi} + \frac{2}{3}\rho_{1/3} \tag{S57}$$

Therefore the effective cyclotron frequency

$$\omega_c = \frac{2\pi}{m_{1/3}}\left|\frac{2}{3}\rho_{1/3} + \frac{1}{3}\frac{B}{2\pi}\right| \tag{S58}$$

where $m_{1/3}$ is the effective mass of $\psi_{1/3}$. On the other hand, the density is given by deviation the filling fraction of electrons from $\nu = 2/3$:

$$\rho_{1/3}\mathcal{A}_{\mathrm{UC}} = 3\delta\nu - 2\delta\nu_B \tag{S59}$$

Using a similar parametrization as the previoius case,

$$\omega_c\tau = \tilde{\tau}\,|2\delta\nu - \delta\nu_B| \tag{S60}$$

where $\tilde{\tau} \equiv \frac{2\pi\tau}{m_{1/3}\mathcal{A}_{\mathrm{UC}}}$ is a dimensionless measure of the disorder strength.

The transition point out of the RIQH regime, by our theory, is determined by when the 0-th (or 1-st) de-localized state crosses the Fermi level. Using Eq. S38, we find these two boundaries should be determined by setting $n = 0, 1$ in the below equation

$$\tilde{\tau}^2\,(3\delta\nu - 2\delta\nu_B)\,(2\delta\nu - \delta\nu_B) = \left(n + \frac{1}{2}\right)\left[1 + (2\delta\nu - \delta\nu_B)^2\,\tilde{\tau}^2\right] \tag{S61}$$

It should be noted that, in reality, the FQAH state spontaneously breaks the time reversal symmetry and we have implicitly assumed that the $B = 0$, $\nu = 2/3$ state has $2\pi\sigma_H = 2/3$. A *positive* $B$ thus flips the Chern number of the parent state, and a time-reversed discussion should be applied in that case. Therefore, the current discussion is only relevant for $B < 0$.

In the main text, we plot these phase boundaries for a particular set of parameters: $\frac{2\pi\tau}{m\mathcal{A}_{\mathrm{UC}}} = 20$ for 1/3 anyons and $\frac{\pi\tau}{m\mathcal{A}_{\mathrm{UC}}} = 40$ for 2/3 anyons. Note that, $\tau_{1/3} = m_{1/3}n_{\mathrm{imp}}(u/3)^2$ whereas $\tau_{2/3} = m_{2/3}n_{\mathrm{imp}}(2u/3)^2$ by our assumptions. This means that $4m_{2/3} = m_{1/3}$ for the chosen parameters.

---



\end{document}